\documentclass[lettersize,journal]{IEEEtran}
\usepackage{amsmath,amsfonts,amssymb}
\usepackage{algorithmic}
\usepackage{algorithm}
\usepackage{array}
\usepackage[caption=false,font=normalsize,labelfont=sf,textfont=sf]{subfig}
\usepackage{textcomp}
\usepackage{stfloats}
\usepackage{url}
\usepackage{verbatim}
\usepackage{graphicx}
\usepackage{cite}
\usepackage{booktabs}
\usepackage{balance}
\usepackage{enumitem}
\graphicspath{{figures/}}
\newcommand{\sys}{vLLM}

\newcommand{\dd}{Cohen's $d$}
\begin{document}

\title{Characterizing Contention-Induced Reliability Collapse in KV-Cache Timing Side Channels for Multi-Tenant LLM Serving}

\author{Rana Abu Bakar%
\thanks{This work has been submitted to IEEE Transactions on Dependable and Secure Computing (TDSC) for possible publication.}
}

\markboth{IEEE Transactions on Dependable and Secure Computing}%
{Rana Abu Bakar: Contention-Induced Reliability Collapse in KV-Cache Timing Side Channels}

\maketitle

\begin{abstract}
Shared key--value (KV) cache reuse improves the efficiency of large language model (LLM) serving, but it can also create a timing side channel that shows whether a prefix is already cached. Previous work has shown that this type of attack is possible, but it is still not clear how reliable the timing signal remains when the server is used by several tenants at the same time. In this paper, we study this problem on live shared LLM-serving systems through seven experiments. We vary background load, request size, measured request overlap, concurrency depth, deployment setup, and serving framework. On the main vLLM server running DeepSeek-R1-Distill-Llama-8B on NVIDIA GB10, the timing signal drops strongly when contention starts. Mean Cohen's $d$ decreases from 0.7789 with no synthetic workers to 0.2109 with two workers ($t=8.412$), while adding more workers does not cause a statistically detectable further drop. A 120-run sparse-overlap experiment gives a breakpoint at the edge of the measured range ($\hat{\tau}=0$, 95\% CI $[0.000,0.113]$), which supports an ambient-versus-loaded regime change instead of an internal physical threshold. AUROC decreases from 0.650 in the ambient condition to 0.531 near 61\% measured overlap and then increases partly to 0.574 at saturation. Concurrency-depth variance has the strongest measured relation with effect size ($r=-0.416$) and hit consistency ($r=-0.637$), which suggests that changes in concurrent activity matter more than overlap level alone. An interleaved experiment keeps the same non-monotonic ordering and reduces the concern about time-of-day effects. Finally, the main collapse is also reproduced on a real two-node, two-GPU tensor-parallel vLLM setup, where mean $d$ falls from 3.418 to 0.511 (Welch $t\approx4.09$, $p<0.01$). Two SGLang pilots do not show the same direction, but both results are statistically inconclusive and were collected under an MPS co-resident setup. Overall, the results show that KV-cache timing reliability depends strongly on the load regime and serving stack, and measurements collected on a quiet system can overestimate the reliability of an attack in a real shared environment.
\end{abstract}


\section{Introduction}
\IEEEPARstart{M}{odern} LLM services multiplex requests from independent users on shared accelerators. Continuous batching, paged KV-cache management, and prefix reuse improve throughput, but they also make latency depend on shared state. Automatic prefix caching is security relevant because a request whose prefix is already cached avoids part of the prefill computation, creating a remotely observable timing difference.

PROMPTPEEK~\cite{wu2025promptpeek}, EarlyBird~\cite{song2024earlybird}, and InputSnatch~\cite{zheng2024inputsnatch} show that cache-timing leakage can support cache-state inference and prompt reconstruction. PROMPTPEEK also studies concurrency, reporting a clear reduction in extracted prompt length beyond roughly 200 configured concurrent users as GPU memory pressure accelerates KV-cache eviction. That result is important, but it measures attack extraction completeness under a configured user-count sweep rather than the statistical reliability of the timing signal itself. SafeKV~\cite{chu2026safekv} and KVGov~\cite{addagada2026kvgov} instead focus on defense and governance. Thus, prior work establishes both feasibility and load sensitivity, but does not characterize effect size, AUROC, hit consistency, or repeated-probe reliability against directly measured contention and concurrency structure.

We study that gap on a live production \sys{} server with uncontrolled real-user traffic present throughout. Synthetic co-tenant traffic is superimposed to vary contention in a repeatable way. The central result is a regime change: the largest loss occurs when sustained synthetic contention appears, not as a smooth function of worker count. We then replace configured worker count with measured request overlap, deliberately sample sparse overlap across the full range, quantify attacker success, and derive concurrency depth from request intervals.

The paper makes five contributions. First, it provides a controlled characterization of KV-cache timing reliability under concurrent serving on a live multi-user stack. Second, it demonstrates a reproducible contention-onset collapse across several workload designs, with a strong zero-to-loaded transition and no detectable monotonic degradation at higher worker counts. Third, it directly tests the threshold hypothesis with 120 sparse-overlap runs and shows that the best breakpoint is at the boundary, while attack-level AUROC and repeated-probe success remain measurable. Fourth, it shows that concurrency-depth variance explains residual instability better than overlap fraction or binary contention presence and validates the main non-monotonic ordering with an interleaved control. Fifth, it adds an independent-node and cross-framework/multi-GPU replication: the central collapse reproduces significantly on distributed two-GPU vLLM, while SGLang pilots remain inconclusive under an MPS co-residency confound.

We do \emph{not} claim that contention is a defense, that the timing channel disappears, or that the observed response is universal across hardware or serving frameworks. Our claim is narrower: operational attack reliability on vLLM is strongly load-regime dependent, simple overlap/utilization proxies are insufficient, and the framework itself may materially shape the observed response.

\section{Background and Related Work}
\subsection{KV-cache timing leakage}
Transformer serving retains attention keys and values for processed tokens. Prefix caching extends reuse across requests: if a new request shares a cached prefix, the server can skip corresponding prefill work. In \sys{}, prefix caching is implemented over cache blocks and prefix context~\cite{vllmapc2026}. Let $L_m$ and $L_h$ denote miss and hit latency. The useful timing signal depends not only on the mean shift
\begin{equation}
\Delta=\mathbb{E}[L_m]-\mathbb{E}[L_h],
\end{equation}
but also on variance. We summarize separation with
\begin{equation}
d=\frac{\bar{x}_m-\bar{x}_h}{s_p},
\end{equation}
where $s_p$ is the pooled variability term used by the probe harness. Values near zero indicate heavy overlap; sign reversals are retained because they reflect instability rather than being clipped away. PagedAttention~\cite{kwon2023pagedattention} and continuous batching make this signal especially load dependent because queueing, active-sequence count, batch admission, and prefill/decode interleaving can change between otherwise identical probes.

\subsection{Prior attacks}
PROMPTPEEK~\cite{wu2025promptpeek} reconstructs prompts with an incremental token-by-token timing loop and explicitly sweeps concurrent-user count up to 1,000. It reports a knee near 200 users: below that point extracted prompt length remains roughly stable, while above it extraction falls as GPU memory pressure causes earlier KV-cache eviction. This is a real concurrency-dependent result, but the dependent variable is extracted prompt length. PROMPTPEEK does not report effect size, AUROC, hit consistency, directly measured request overlap, concurrency depth, or a change-point test on timing-signal distinguishability.

EarlyBird~\cite{song2024earlybird} validates timing attacks on open-source and commercial services; its reported classifier operating point reaches TPR 0.99 and FPR 0.003. It also notes GPU-system noise and power fluctuations as factors that weaken classification, but treats them as noise to mitigate rather than as a controlled contention variable. InputSnatch~\cite{zheng2024inputsnatch} combines candidate construction with statistical timing analysis, reporting 87.13\% cache-hit-prefix-length accuracy, 62\% exact extraction in its medical setting, and 43--100\% success across legal-query types. It explicitly notes temporal noise from concurrent request processing, but filters that noise rather than measuring attack reliability as the concurrent load changes.

\subsection{Defenses and serving systems}
SafeKV~\cite{chu2026safekv} selectively isolates sensitive cache entries and reports an average defense success rate above 94\%, together with performance improvements such as TTFT overhead reduction from 50.41\% to 11.74\% on Qwen-235B-A22B and throughput gains up to $2.66\times$ against a partitioned-cache baseline. These are defense-efficacy and cost metrics, not residual attacker-reliability curves under varied load.

KVGov~\cite{addagada2026kvgov} cryptographically separates cache identities across tenants and confirms strong cold-versus-cached timing ratios on real hardware (0.22 on vLLM/A100 and 0.093 in an independent llama.cpp/Metal replication). Its audit-scheduler utility analysis is simulation based, and the paper explicitly leaves extended field experiments with real adversarial tenants as future work. It also contrasts fixed simulation latencies with the variable latency caused by GPU scheduling, memory-bandwidth contention, and network jitter in real deployments.

Our work therefore does not claim that prior studies ignore concurrency. The narrower gap is that none of these systems measures the \emph{timing signal's own statistical reliability} as a function of directly measured overlap or concurrency structure, tests whether the reliability change is an interior threshold or an ambient-versus-loaded split, and then rechecks the result against a time-of-day confound on an always-on production server. Studies 1--7 are designed around these questions, including an explicit cross-deployment replication.

\section{Research Questions}
We organize the evaluation around six questions. \textbf{RQ1} asks whether sustained co-tenant activity reduces cache-timing distinguishability. \textbf{RQ2} asks whether degradation is gradual with configured load or dominated by a regime transition. \textbf{RQ3} tests whether the effect is specific to one synthetic workload shape. \textbf{RQ4} replaces configured worker count with directly measured temporal overlap. \textbf{RQ5} samples the full overlap range to test for an interior breakpoint and asks whether concurrency depth explains residual variance better than overlap fraction. \textbf{RQ6} asks whether the central result survives an independent physical instance and genuine two-GPU distributed serving, and whether it transfers from vLLM to SGLang.

\begin{table*}[t]
\centering
\caption{Experimental progression. Each study addresses a limitation exposed by the previous one rather than repeating the same load sweep.}
\label{tab:study-overview}
\footnotesize
\setlength{\tabcolsep}{3pt}
\begin{tabular}{@{}p{0.055\linewidth}p{0.075\linewidth}p{0.125\linewidth}p{0.18\linewidth}p{0.43\linewidth}@{}}
\toprule
Study & Runs & Controlled factor & Primary measurement & Main purpose / result \\
\midrule
1 & 10 & Ambient vs. 4 workers & $d$, $H$ & Pilot establishes a large first-order loss when sustained synthetic contention is introduced. \\
2 & 48 & 0/2/4/8 workers & $d$, $H$, Welch tests & Dose response shows the dominant transition is 0$\rightarrow$2 workers; higher worker counts add no statistically detectable monotonic loss. \\
3 & 36 & Worker count and output length & $d$, $H$ & Disentangles filler shape and exposes the unusually high variance of one short-output worker. \\
4 & 20 & Workload shape with interval logging & $O$, $d$, $H$ & Shows request-overlap fraction rapidly saturates and cannot explain residual loaded-regime variance. \\
5 & 120 & Sparse measured overlap across $[0,1]$ & AUROC, $H$, $P_{succ}$, $C(t)$, breakpoint & Tests the missing transition region, rejects an interior threshold, and identifies concurrency-depth variance as the strongest measured correlate. \\
6 & 45 & Interleaved 0/40/100\% conditions & $d$ vs. round/order & Rechecks the non-monotonic ordering while reducing long-timescale time-of-day confounding. \\
\bottomrule
\end{tabular}
\end{table*}

\section{Threat Model and Experimental Method}
\subsection{Threat model}
The adversary is a remote tenant able to submit requests and measure response time. It has no privileged server, GPU, driver, or scheduler telemetry. It can issue controlled before/after probes around a candidate prefix and repeat them. We do not recover real users' prompts; all cache-state discrimination uses controlled probe material. Other users may submit unrelated traffic throughout the campaign.

\subsection{Serving platform}
The primary target is a live \sys{} 0.28.0 server running \texttt{deepseek-ai/DeepSeek-R1-Distill-Llama-8B} on one NVIDIA GB10 (DGX Spark). The environment uses PyTorch 2.13.0+cu130, CUDA 13.0, and NVIDIA driver 580.95.05; the launch configuration is \texttt{--dtype half --gpu-memory-utilization 0.70 --max-model-len 4096 --enforce-eager}. Prefix caching was verified from live \texttt{/metrics} counters~\cite{vllmmetrics2026} (1,393,313 queried and 568,208 hit tokens). Real production traffic was never suppressed, so the zero-synthetic-worker condition is \emph{ambient}, not idle. Study~7 adds a second physical GB10 node and two-node tensor-parallel deployments of vLLM 0.26.0 and SGLang 0.5.17 over NCCL; these replications are reported separately because their ambient conditions are not all equivalent.

The probe scripts record paired uncached/cached timings. Probe completions are non-streaming with \texttt{max\_tokens=1}; elapsed time uses \texttt{time.perf\_counter()}, while wall-clock intervals for overlap use \texttt{time.time()}. Synthetic load is generated by configurable workers and generation length; an instrumented version records each background interval $I_j=[s_j,e_j]$.

\subsection{Metrics}
Hit consistency is
\begin{equation}
H=\frac{1}{N}\sum_{i=1}^{N}\mathbf{1}[\text{pair }i\text{ has expected ordering}],
\end{equation}
and is reported directly from raw runs with bootstrap confidence intervals. For binary detection, ``before'' probes are uncached negatives and ``after'' probes cached positives. AUROC is computed with the Mann--Whitney rank estimator
\begin{equation}
\mathrm{AUROC}=\frac{R_+-n_+(n_++1)/2}{n_+n_-},
\end{equation}
where $R_+$ is the positive-class rank sum. Accuracy, balanced accuracy, TPR, and FPR use the Youden-optimal latency threshold. AUROC confidence intervals use bootstrap resampling. Consecutive worker levels in Study~2 are compared with Welch's $t$ test because the observed variances differ; degrees of freedom use the Welch--Satterthwaite approximation.

For odd repeated-probe budgets $k\in\{1,3,5,9\}$, majority-vote success is
\begin{equation}
P_{\mathrm{succ}}(k)=\sum_{i=\lfloor k/2\rfloor+1}^{k}{k\choose i}p^i(1-p)^{k-i},
\end{equation}
where $p$ is the empirical single-pair hit rate.

\label{sec:overlap}
For a probe interval $P_i$ and union of synthetic intervals $B=\cup_j I_j$, measured request overlap is
\begin{equation}
O_i=\frac{|P_i\cap B|}{|P_i|}.
\end{equation}
This is a request-layer temporal overlap fraction, not GPU SM occupancy. \label{sec:concurrency-depth-metric}
We also derive instantaneous synthetic concurrency depth
\begin{equation}
C(t)=\sum_j\mathbf{1}[s_j\le t<e_j],
\end{equation}
with per-probe mean, maximum, and variance. These retain information discarded by the binary predicate $\mathbf{1}[C(t)>0]$.

\subsection{Change-point model}
\label{sec:changepoint-method}
Study~5 fits two \emph{independent} linear segments,
\begin{equation}
d(O)=\begin{cases}\alpha_1+\beta_1O,&O\le\tau,\\\alpha_2+\beta_2O,&O>\tau,\end{cases}
\end{equation}
with no continuity constraint. We grid-search observed $O$, require at least four points per segment, and select the minimum-SSE split. The breakpoint and slopes use 2,000 bootstrap refits. We do not claim an interior physical threshold unless the breakpoint CI lies inside the observed range.

\section{Study 1: Pilot}
\label{sec:study1}

\subsection{Objective}

The pilot asked whether introducing a small sustained synthetic workload
produces any observable change in the cache-timing signal.

Five independent runs were collected per condition.

\subsection{Results}

\begin{table}[t]
\centering
\caption{Study 1: pilot effect-size and hit-consistency summary, with
bootstrap 95\% CIs (2,000 resamples).}
\label{tab:pilot}
\begin{tabular}{lccc}
\toprule
Condition & Mean $d$ [95\% CI] & Std. dev. & Mean $H$ [95\% CI] \\
\midrule
Ambient & 1.126 [0.923, 1.329] & 0.269 & 88.0\% [83.2, 92.8] \\
$+4$ workers & 0.213 [0.152, 0.298] & 0.096 & 63.2\% [53.6, 72.8] \\
\bottomrule
\end{tabular}
\end{table}

The mean effect size decreases from 1.126 to 0.213 when four synthetic
workers are introduced.

The absolute change is

\begin{equation}
\Delta d = 1.126 - 0.213 = 0.913.
\end{equation}

Expressed relative to the ambient mean, the loaded condition retains
only

\begin{equation}
\frac{0.213}{1.126} \approx 18.9\%
\end{equation}

of the pilot's standardized effect.

\subsection{Interpretation}

The pilot strongly suggests that concurrent activity can destabilize the
timing channel, but it cannot reveal the response shape. In particular,
it does not tell us whether the effect decreases smoothly with worker
count or whether most of the change occurs at contention onset.

This motivates Study~2.

\section{Study 2: Worker-Count Dose Response}
\label{sec:study2}

\subsection{Design}

Study~2 fixes the synthetic request size and varies only worker count:

\begin{equation}
W \in \{0,2,4,8\}.
\end{equation}

Each condition contains 12 independent runs.

\subsection{Effect-Size Results}

\begin{table}[t]
\centering
\caption{Study 2: effect size and hit consistency versus configured
worker count, with bootstrap 95\% CIs (2,000 resamples, $n=12$ per
level).}
\label{tab:dose}
\begin{tabular}{rccc}
\toprule
Workers & Mean $d$ [95\% CI] & Std. dev. & Mean $H$ [95\% CI] \\
\midrule
0 & 0.7789 [0.713, 0.845] & 0.1241 & 79.0\% [78.0, 80.2] \\
2 & 0.2109 [0.107, 0.318] & 0.1982 & 63.3\% [57.7, 68.8] \\
4 & 0.1399 [0.016, 0.262] & 0.2250 & 61.0\% [53.8, 67.7] \\
8 & 0.0820 [$-$0.028, 0.201] & 0.2093 & 58.7\% [49.3, 67.7] \\
\bottomrule
\end{tabular}
\end{table}

The largest change occurs at the first loaded condition:

\begin{equation}
0.7789 \rightarrow 0.2109.
\end{equation}

The later means continue to decrease numerically, but the changes are
much smaller relative to the within-condition variance.

The relative effect retained at two workers is

\begin{equation}
\frac{0.2109}{0.7789} \approx 27.1\%.
\end{equation}

At eight workers it is

\begin{equation}
\frac{0.0820}{0.7789} \approx 10.5\%.
\end{equation}

These ratios are descriptive only; inferential conclusions are based on
the run-level comparisons.

\subsection{Welch Tests}

\begin{table}[t]
\centering
\caption{Study 2: Welch tests between consecutive worker levels.}
\label{tab:welch}
\begin{tabular}{lcc}
\toprule
Comparison & $t$ & Welch df \\
\midrule
0 vs.\ 2 workers & 8.412 & 18.5 \\
2 vs.\ 4 workers & 0.820 & 21.7 \\
4 vs.\ 8 workers & 0.652 & 21.9 \\
\bottomrule
\end{tabular}
\end{table}

The zero-to-two-worker transition is large relative to the sampling
variance. By contrast, the two later transitions are not statistically
distinguishable at conventional significance levels.

This is the key result of the paper.

\begin{figure}[t]
\centering
\includegraphics[width=\linewidth]{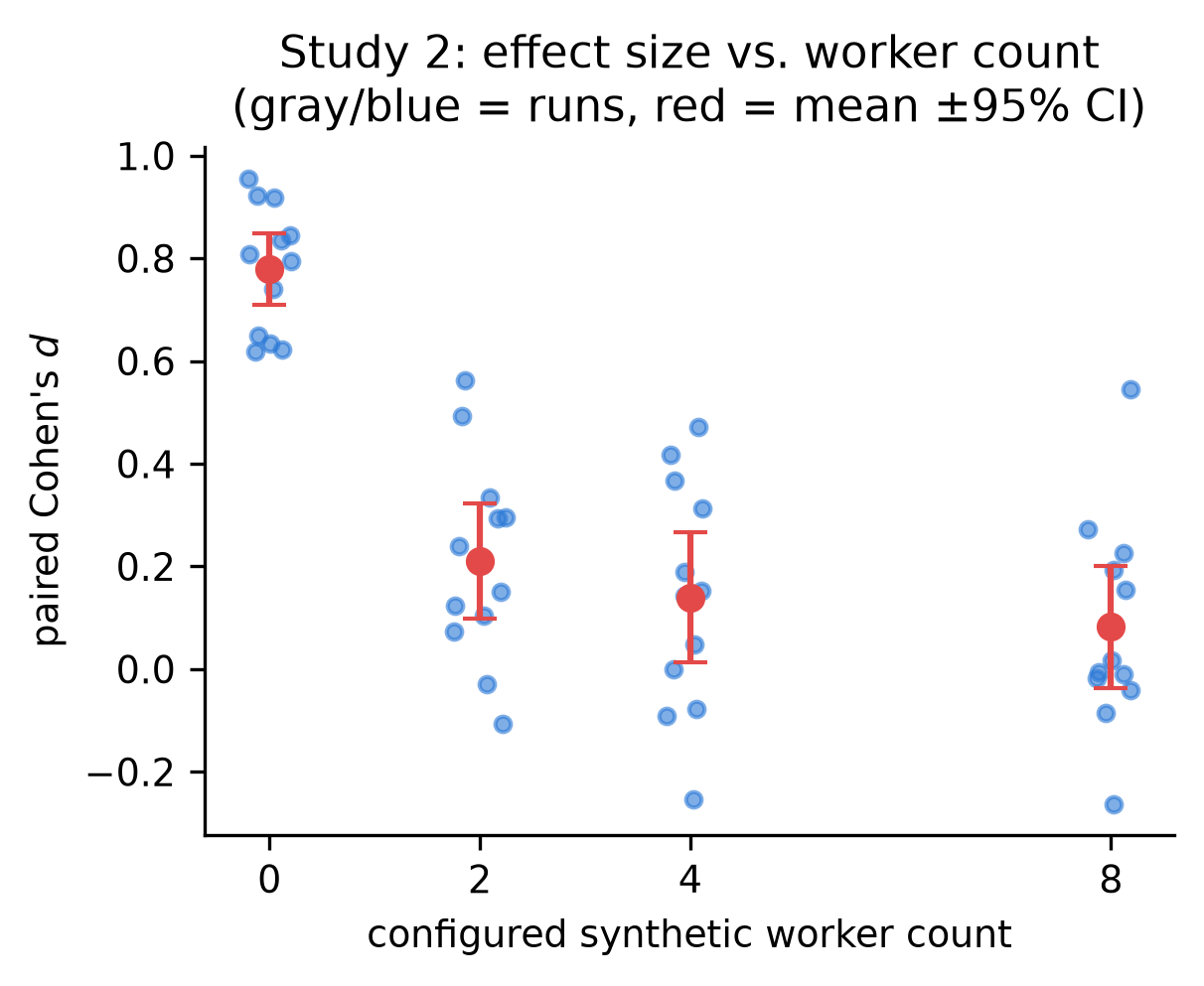}
\caption{Study 2: every run plotted as a jittered point against
configured worker count, with the condition mean and bootstrap 95\% CI
in red. The near-total drop occurs between 0 and 2 workers; the
2/4/8-worker means remain within one another's CIs.}
\label{fig:study2}
\end{figure}

\subsection{Contention-Onset Regime Change Rather Than Linear Degradation}

A naive linear-load hypothesis predicts a progressively weaker timing
signal as worker count rises. The data do not support that qualitative
pattern as the dominant effect.

Instead, the evidence is better summarized as

\begin{equation}
\text{ambient regime}
\rightarrow \text{collapsed loaded regime}.
\end{equation}

Study~2 alone does not localize a physical threshold because worker count is only a configuration knob. We therefore describe this result as a \emph{contention-onset transition}. Study~5 (Section~\ref{sec:study5}) tests this distinction
directly on a fine-grained, measured-overlap variable rather than worker
count, and its bootstrapped breakpoint estimate confirms the same
conclusion: the data support an ambient-versus-loaded regime change, not
an interior physical threshold.

\subsection{Why Worker Count Is Not a Physical Load Metric}

Worker count changes the number of independently issuing clients, but
its relationship to instantaneous GPU work is mediated by request
duration, server queueing, continuous batching, and ambient traffic.

For example, two workers generating long outputs may keep the server
continuously busy, while a larger number of workers issuing very short
requests may create different batch dynamics.

This motivates Study~3.

\section{Study 3: Disentangling Parallelism and Request Size}
\label{sec:study3}

\subsection{Design}

Study~3 changes worker count and maximum generation length
independently. Three configurations are evaluated with 12 runs each.

\begin{table*}[t]
\centering
\caption{Study 3: independently varied filler configurations, with hit
consistency and bootstrap 95\% CIs (2,000 resamples, $n=12$ per
condition).}
\label{tab:disentangle}
\begin{tabular}{lrrcccc}
\toprule
Condition & Workers & Max tok. & Mean $d$ [95\% CI] & Std. $d$ & Mean $H$ [95\% CI] \\
\midrule
1w\_small16 & 1 & 16 & 0.0568 [$-$0.317, 0.448] & 0.7112 & 57.8\% [46.7, 69.2] \\
1w\_big128 & 1 & 128 & 0.0854 [0.023, 0.156] & 0.1207 & 62.3\% [57.5, 66.7] \\
2w\_tiny4 & 2 & 4 & 0.1089 [0.068, 0.147] & 0.0727 & 61.8\% [60.3, 63.5] \\
\bottomrule
\end{tabular}
\end{table*}

\subsection{Common Low-Mean Regime}

Despite substantial differences in workload shape, all three
configurations produce low mean effect sizes:

\begin{equation}
0.0568 \le \bar{d} \le 0.1089.
\end{equation}

This narrow mean band is consistent with the collapsed regime observed
in Study~2.

The result weakens a simple alternative explanation in which only one
particular worker count or token limit accidentally interferes with the
probe.

\subsection{The Single-Fast-Worker Anomaly}

The most interesting difference is not in the mean but in variance.

For \texttt{1w\_small16},

\begin{equation}
s_d = 0.7112,
\end{equation}

whereas for \texttt{2w\_tiny4},

\begin{equation}
s_d = 0.0727.
\end{equation}

Thus, the single-fast-worker condition is nearly an order of magnitude
more variable by this summary statistic.

This matters because a low average effect can arise in two qualitatively
different ways:

\begin{enumerate}[leftmargin=*]
\item every run produces a consistently weak timing signal; or
\item individual runs alternate between strong positive, weak, and reversed
signals, producing a low average after cancellation.
\end{enumerate}

The single-fast-worker condition is closer to the second pattern.

\subsection{Mechanistic Hypothesis}

We hypothesize that this variance is driven by scheduler alignment or
instantaneous concurrency depth.

One short-output worker can repeatedly alternate between active and
inactive states at a cadence comparable to the probe sequence. Depending
on alignment, a cached or uncached member of a pair may see different
queue states.

Two continuously issuing workers can instead maintain a more stable
loaded state even if their individual requests are shorter.

This explanation is plausible but not established by Study~3 alone. It
is therefore treated as a hypothesis and tested only partially through
interval instrumentation in Study~4.

\section{Study 4: Direct Temporal-Overlap Measurement}
\label{sec:study4}

\subsection{Motivation}

Studies~1--3 infer contention from configured worker count and request
size. Study~4 instruments request intervals so that overlap with the
probe can be measured directly.

Each condition contains four runs.

\subsection{Measured Summary}

\begin{table*}[t]
\centering
\caption{Study 4: measured synthetic request-overlap fraction, effect-size
range, and mean effect size / hit consistency with bootstrap 95\% CIs
(2,000 resamples, $n=4$ per condition). ``Ambient'' means no synthetic
workers; uncontrolled production traffic may still be present.}
\label{tab:occupancy}
\begin{tabular}{lcccc}
\toprule
Condition & Measured synthetic overlap & $d$ range & Mean $d$ [95\% CI] & Mean $H$ [95\% CI] \\
\midrule
ambient & 0.0000 in all four runs & 0.34--0.79 & 0.496 [0.364, 0.700] & 75.5\% [73.0, 78.5] \\
1w\_small16 & 0.965--0.997 & $-0.91$--0.79 & 0.124 [$-$0.551, 0.658] & 53.5\% [36.5, 70.0] \\
2w\_16 & 0.970--1.000 & 0.17--0.41 & 0.281 [0.174, 0.388] & 73.5\% [69.0, 78.0] \\
1w\_big128 & 0.760--0.930 & $-0.11$--0.22 & 0.099 [$-$0.049, 0.202] & 69.5\% [67.0, 72.5] \\
2w\_tiny4 & 0.990--1.000 & 0.12--0.18 & 0.151 [0.131, 0.168] & 67.5\% [60.5, 74.0] \\
\bottomrule
\end{tabular}
\end{table*}

\subsection{Overlap Saturation}

The instrumentation confirms that sustained synthetic workloads overlap
the probe for most of its wall-clock lifetime. Even a single
small-request worker produces

\begin{equation}
O \approx 0.965\text{--}0.997.
\end{equation}

The two-worker tiny-request condition produces approximately

\begin{equation}
O \approx 0.990\text{--}1.000.
\end{equation}

This saturation is itself an important result because it explains why
overlap fraction has little dynamic range among the loaded
configurations.

\subsection{Correlation with Effect Size}

Across all 20 Study~4 runs,

\begin{equation}
r_{O,d}=-0.3662.
\end{equation}

The sign is consistent with the broad hypothesis that more contention
reduces the timing signal.

However, the magnitude is not evidence for a clean linear relationship.
Most of the observable association is generated by separation between
the ambient cluster at $O=0$ and the loaded cluster at high $O$.

Within the loaded cluster, overlap provides little ability to rank
effect-size reliability.

\begin{figure}[t]
\centering
\includegraphics[width=\linewidth]{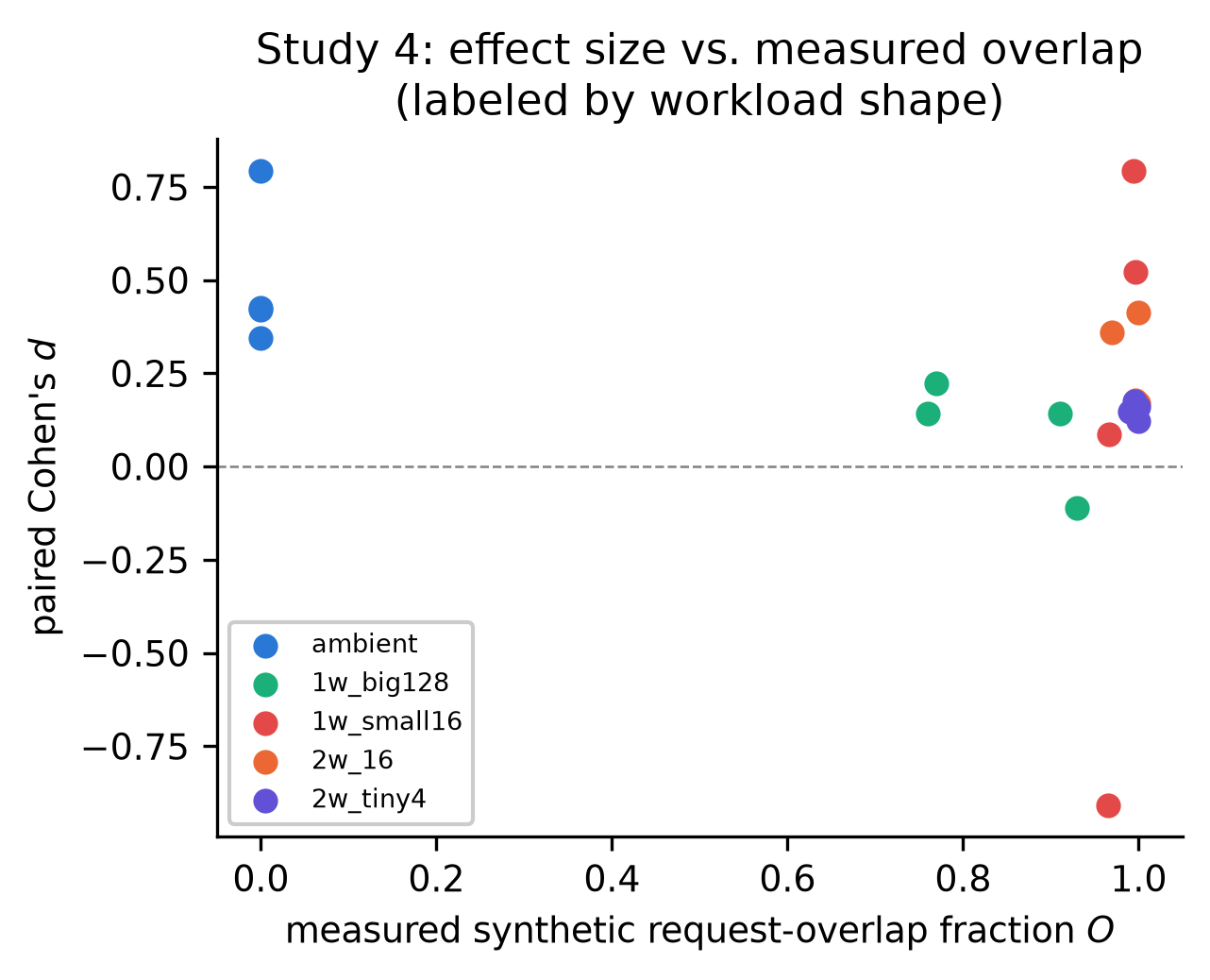}
\caption{Study 4: effect size versus measured overlap, labeled by
workload shape. The loaded cluster sits at $O\gtrsim0.76$ regardless of
workload, but \texttt{1w\_small16} (red) spans nearly the full range of
observed $d$, from the single most negative point to among the most
positive, despite similar overlap to the other loaded conditions.}
\label{fig:study4-labeled}
\end{figure}

\subsection{Why ``GPU Occupancy'' Would Be Incorrect}

The recorded intervals begin and end at the request layer. They do not
expose CUDA kernel residency or SM active cycles.

Accordingly,

\begin{equation}
O \neq \text{GPU occupancy}.
\end{equation}

The metric is also distinct from \sys{} KV-cache utilization and from
the number of requests currently running.

A production implementation can observe server-side quantities such as
running-request count, KV-cache usage, prefix-cache query/hit counters,
TTFT histograms, and prompt/generation token counters through \sys{}
metrics, depending on the deployed version.

Those metrics should be incorporated in a follow-on experiment.

\subsection{Failure to Explain the Single-Worker Variance}

Study~4 provides a particularly informative comparison.

Both \texttt{1w\_small16} and \texttt{2w\_tiny4} spend nearly all probe
time overlapped with at least one synthetic request.

Yet their timing behavior differs sharply. The former can range from a
large negative effect to a substantial positive effect, while the latter
remains consistently near the collapsed band.

A binary overlap predicate cannot represent this distinction because
both conditions map almost entirely to

\begin{equation}
\mathbf{1}[C(t)>0] = 1.
\end{equation}

This is direct evidence that request-overlap fraction is an insufficient
second-order explanatory variable.

\section{Study 5: Sparse-Overlap Sampling, Attack Reliability, and Concurrency Depth}
\label{sec:study5}

\subsection{Motivation}

Study~4 left two questions explicitly open: whether the collapse is an
interior physical threshold in overlap or an ambient-versus-loaded
regime change, and whether concurrency depth explains more residual
variance than the binary/fractional overlap metric. This study answers
both directly, using only new scripts and without modifying
\texttt{live\_short\_instrumented.py}, \texttt{background\_load\_v3.py},
or \texttt{compute\_occupancy.py}.

\subsection{Design}

A burst/sleep generator, \path{sparse_contention_load.py}, alternates
idle and active phases on a fixed wall-clock cycle to target a
configured duty cycle. Because achieved overlap depends on request
latency as well as configuration, we first calibrate: 22 candidate
configurations (1 or 2 workers, duty-cycle targets from 0.02 to 1.0,
6-second cycle, 8-token filler completions) are each run once against a
live probe, and the configuration whose measured overlap is closest to
each of eight target bins,

\begin{equation}
O \in \{0,0.05,0.10,0.20,0.40,0.60,0.80,1.00\},
\end{equation}

is selected. Fifteen independent runs are then collected per bin (120
runs total), each analyzed using its own \emph{measured} overlap
(Section~\ref{sec:overlap}), never the configured target. The ambient
bin ($O=0$) uses no synthetic generator at all.

\subsection{Effect-Size and Attack-Reliability Results}

\begin{table*}[t]
\centering
\caption{Study 5: pooled attack reliability per bin (750 trials per bin, 15 runs $\times$ 50 trials).}
\label{tab:study5reliability}
\begin{tabular}{lcccccccc}
\toprule
Bin & Mean $O$ & AUROC & 95\% CI & $p_{\mathrm{hit}}$ & $P_{\mathrm{succ}}(1)$ & $P_{\mathrm{succ}}(3)$ & $P_{\mathrm{succ}}(5)$ & $P_{\mathrm{succ}}(9)$ \\
\midrule
0\%   & 0.000 & 0.650 & [0.628, 0.675] & 0.695 & 0.695 & 0.777 & 0.830 & 0.895 \\
5\%   & 0.076 & 0.619 & [0.595, 0.644] & 0.648 & 0.648 & 0.716 & 0.762 & 0.825 \\
10\%  & 0.096 & 0.628 & [0.602, 0.652] & 0.652 & 0.652 & 0.721 & 0.768 & 0.832 \\
20\%  & 0.184 & 0.586 & [0.564, 0.611] & 0.639 & 0.639 & 0.703 & 0.747 & 0.809 \\
40\%  & 0.392 & 0.564 & [0.541, 0.587] & 0.575 & 0.575 & 0.611 & 0.638 & 0.678 \\
60\%  & 0.613 & 0.531 & [0.507, 0.556] & 0.525 & 0.525 & 0.538 & 0.547 & 0.562 \\
80\%  & 0.749 & 0.552 & [0.524, 0.578] & 0.560 & 0.560 & 0.590 & 0.611 & 0.645 \\
100\% & 0.995 & 0.574 & [0.544, 0.603] & 0.584 & 0.584 & 0.625 & 0.655 & 0.699 \\
\bottomrule
\end{tabular}
\end{table*}

AUROC and $p_{\mathrm{hit}}$ decline from ambient through the 60\% bin, then partially recover at 80\% and 100\%; AUROC remains above chance (0.5) in every bin, bottoming at 0.531. This non-monotonic shape was not predicted by Studies~1--4 and is discussed further in Section~\ref{sec:study5-caveat}.

We use odd probe budgets $k\in\{1,3,5,9\}$ so that strict-majority voting
never requires a tie-breaking rule (an even $k$, such as 10, forces an
arbitrary choice for a 50/50 split, and that choice alone -- not the
underlying noise -- can flip whether success probability appears to rise
or fall with $k$). With odd $k$, $P_{\mathrm{succ}}(k)$ increases
monotonically with $k$ in every bin, including the near-chance 60\% bin
(0.525 at $k=1$ to 0.562 at $k=9$), consistent with majority voting
behaving as expected once the tie artifact is removed; the improvement
is modest in the low-reliability bins because it is bounded by how far
the per-trial hit rate sits above 0.5.

\subsection{Concurrency-Depth Analysis}

For every probe window, we derive concurrency depth directly from the
same interval logs used for overlap (Section~\ref{sec:overlap}):
the count of background requests whose interval overlaps the probe's
window, aggregated per run as mean, maximum, and variance
(Section~\ref{sec:concurrency-depth-metric}).

\begin{table}[t]
\centering
\caption{Study 5: correlation of contention metrics with \dd{} and hit consistency (n=120 runs).}
\label{tab:concurrency-corr}
\begin{tabular}{lcc}
\toprule
Metric & $r$ with $d$ & $r$ with $H$ \\
\midrule
Measured overlap $O$ & $-0.106$ & $-0.504$ \\
Mean binary overlap & $-0.122$ & $-0.510$ \\
Mean concurrency depth & $+0.039$ & $-0.337$ \\
Max concurrency depth & $-0.235$ & $-0.419$ \\
Concurrency-depth variance & $\mathbf{-0.416}$ & $\mathbf{-0.637}$ \\
\bottomrule
\end{tabular}
\end{table}

Concurrency-depth variance is the strongest correlate of both \dd{} and
hit consistency, clearly exceeding measured overlap and binary presence.
It remains the strongest correlate ($r=-0.329$) of the \emph{residual}
of \dd{} after regressing on measured overlap alone, indicating that it
captures information the overlap metric discards rather than merely
re-deriving the same signal. This is correlational evidence supporting
the hypothesis raised after Study~3 -- that it is the \emph{burstiness}
of contention, not its average level, that best predicts instability --
but with ambient production load on the shared host uncontrolled
throughout, correlation alone does not establish burstiness as the
causal mechanism; it is the strongest measured correlate found in this
study, not a confirmed cause. Importantly, $C(t)$ is derived only from request-layer wall-clock intervals; no vLLM scheduler-internal event, queue depth, batch-admission decision, or GPU kernel schedule is observed. The result therefore identifies the strongest measured request-layer correlate, not the internal mechanism.

\begin{figure}[t]
\centering
\includegraphics[width=\linewidth]{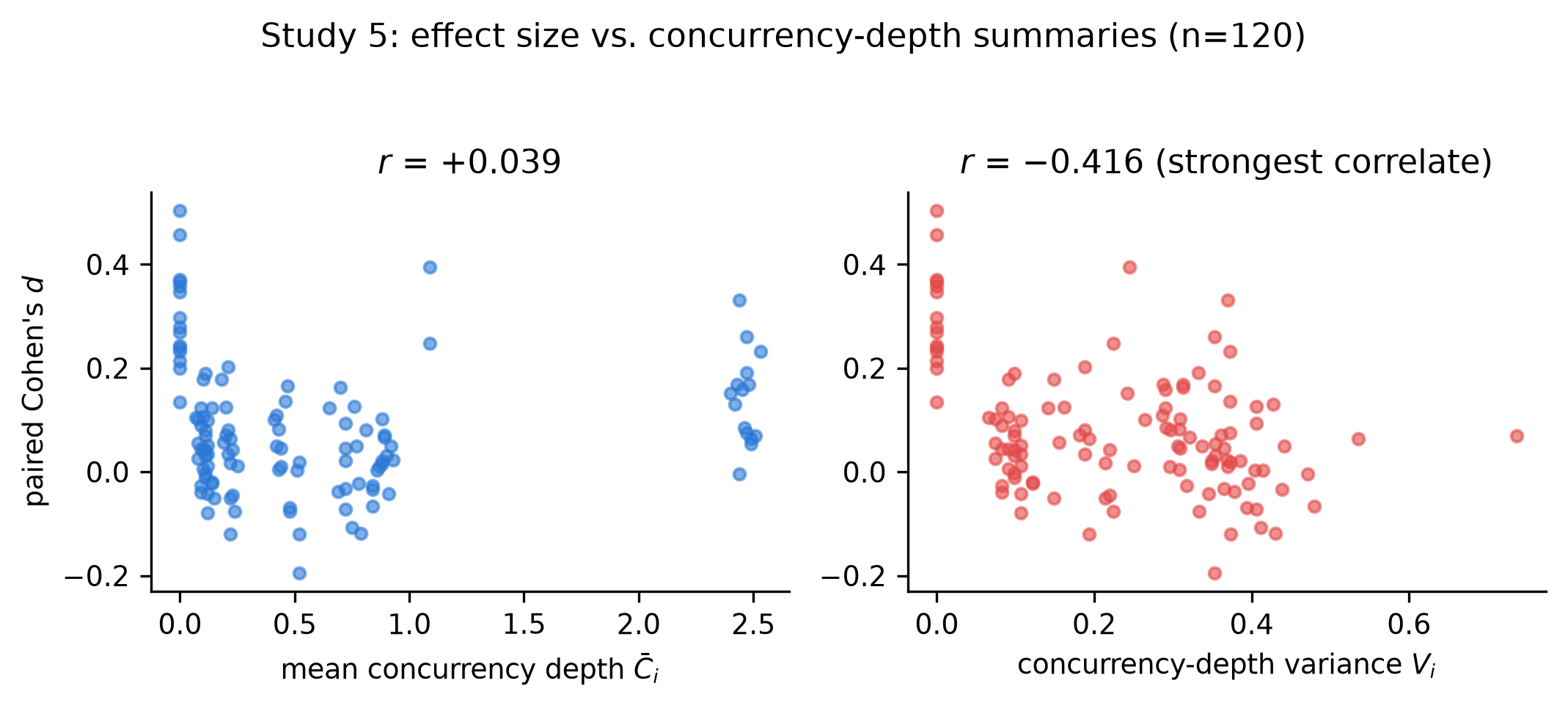}
\caption{Study 5: effect size versus mean concurrency depth (left,
$r=+0.039$) and concurrency-depth variance (right, $r=-0.416$, the
strongest correlate found in this study). Mean depth clusters into
discrete bands set by worker count; variance spreads continuously and
tracks \dd{} far better.}
\label{fig:concurrency-depth}
\end{figure}

\subsection{Segmented Regression: Threshold or Regime Change?}

Applying the breakpoint procedure of Section~\ref{sec:changepoint-method}
to the pooled 120-run $(O,d)$ sample gives a single-line fit of
$R^2=0.011$ versus a two-segment fit of $R^2=0.536$ at
$\hat\tau=0$ (bootstrap 95\% CI $[0.000,0.113]$), with fitted segment
1 (all 15 points at $O=0$, so this is simply their mean, not a
regression in the usual sense) at $d\approx0.301$ and segment 2
($O>\hat\tau$) following $d=0.0171+0.0794\,O$ (slope 95\% CI $[0.015,
0.183]$); the slope difference between segments (95\% CI $[-2.633,
-0.015]$) excludes zero. Because the two segments are fit independently
(Section~\ref{sec:changepoint-method}), a substantial part of this
$R^2$ improvement comes from the discontinuous level shift the
unconstrained fit is free to place at $\hat\tau$ -- segment 2's
intercept ($0.017$) sits well below segment 1's mean ($0.301$) -- rather
than from a smooth change in slope alone; this is itself consistent
with (and additional evidence for) a regime change rather than a
continuously varying physical threshold.

Per the pre-registered decision rule of
Section~\ref{sec:changepoint-method}, this result does \emph{not}
support an interior physical threshold: the breakpoint CI touches the
edge of the observed overlap range rather than sitting in its interior.
It instead reproduces, with a purpose-built experiment, exactly the
ambient-versus-loaded regime distinction already suggested by Studies
1--4 -- and further shows that within the loaded regime, \dd{} rises
slightly rather than staying flat or continuing to fall, consistent
with the AUROC recovery observed at 80--100\% overlap.

\begin{figure}[t]
\centering
\includegraphics[width=\linewidth]{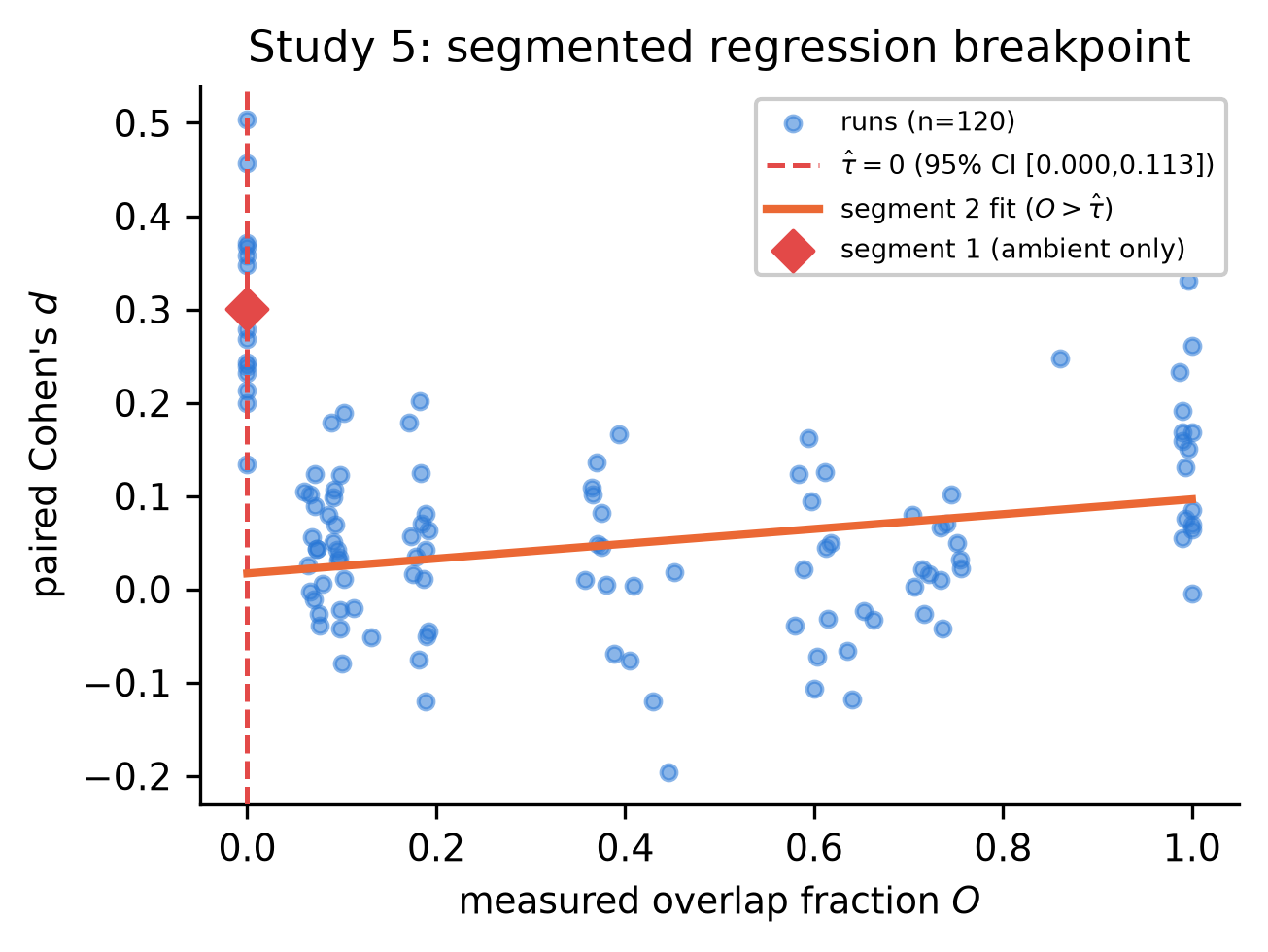}
\caption{Study 5: all 120 runs plotted against measured overlap, with
the fitted breakpoint $\hat\tau=0$ and the segment-2 fit
($O>\hat\tau$). The near-flat, slightly positive segment-2 slope and the
breakpoint sitting at the left edge of the range are visually consistent
with an ambient-vs-loaded regime change rather than an interior
threshold.}
\label{fig:study5-breakpoint}
\end{figure}

\subsection{An Important Environmental Caveat}
\label{sec:study5-caveat}

Study~5 was executed as one continuous $\sim$2.5-hour block later in the
day than Studies~1--4. Its ambient bin shows a substantially weaker
baseline (mean $d\approx0.13$--$0.50$, AUROC 0.650) than the ambient
conditions in Studies~1--4 (mean $d$ up to 1.126). Because real
production traffic on the shared host is not controlled by the
experiment, this may reflect time-of-day drift in ambient load rather
than a property of the attack itself, and it may also explain why the
mid-range dip and high-range partial recovery in Table~\ref{tab:study5reliability}
looks qualitatively different from the flat loaded regime in Study~2.
Study~6 (Section~\ref{sec:study6}) addresses this directly by
interleaving conditions instead of running them as sequential blocks.

\section{Study 6: Interleaved Time-of-Day Drift Control}
\label{sec:study6}

\subsection{Motivation}

Study~5 ran each of its eight bins as one continuous sequential block,
so the mid-range dip and high-overlap partial recovery in
Table~\ref{tab:study5reliability} are confounded with wall-clock time:
if real ambient production load drifted over the $\sim$2.5-hour
campaign, a purely time-driven change could masquerade as a
load-dependent one. Study~6 isolates this by re-running the three most
diagnostic conditions -- ambient (0\%), the mid-range dip (40\%), and
saturation (100\%) -- \emph{interleaved}: each of 15 rounds probes all
three conditions back-to-back, in the same order, before repeating. Any
genuine time-of-day drift should now appear as a trend in \dd{}
\emph{within} a single condition across rounds, cleanly separated from
the between-condition contention effect.

\subsection{Results}

\begin{table}[t]
\centering
\caption{Study 6: interleaved conditions, 15 rounds each. ``corr(d,round)'' and
``slope'' test for a within-condition trend across the interleaved rounds,
i.e.\ time-of-day drift isolated from the contention effect.}
\label{tab:study6}
\begin{tabular}{lccc}
\toprule
Bin & Mean $d$ & Std. $d$ & corr($d$,round) \\
\midrule
0\%   & $+0.2489$ & 0.0778 & $-0.366$ \\
40\%  & $+0.0331$ & 0.0584 & $-0.040$ \\
100\% & $+0.0695$ & 0.0642 & $-0.317$ \\
\bottomrule
\end{tabular}
\end{table}

Two results stand out. First, the ambient and 100\% conditions exhibit
negative temporal trends across rounds (corr($d$,round) $=-0.366$ and
$-0.317$; slopes $-0.0064$ and $-0.0046$ per round respectively), while
the 40\% condition is approximately flat (corr $=-0.040$, slope
$-0.0005$) -- essentially no linear trend. We do not report confidence
intervals or significance tests for these per-condition slopes/
correlations ($n=15$ rounds each), so this observation is reported
descriptively rather than as an established effect: these values are
consistent with modest temporal drift in two of the three conditions,
but the present sample size does not support treating drift as
independently established, and the flat 40\% trend is itself evidence
against a uniform, monotonic drift story affecting all conditions
equally.

Second, and more importantly, the between-condition ordering from
Study~5 \emph{survives} interleaving: $0\%>100\%>40\%$ in both designs.

\begin{table}[t]
\centering
\caption{Study 6 vs.\ Study 5: same three conditions, sequential-block
design (Study~5) versus interleaved design (Study~6).}
\label{tab:study5v6}
\begin{tabular}{lcc}
\toprule
Bin & Study~5 (sequential) $d$ & Study~6 (interleaved) $d$ \\
\midrule
0\%   & $0.3007\pm0.0998$ & $0.2489\pm0.0778$ \\
40\%  & $0.0177\pm0.0995$ & $0.0331\pm0.0584$ \\
100\% & $0.1424\pm0.0888$ & $0.0695\pm0.0642$ \\
\bottomrule
\end{tabular}
\end{table}

The magnitudes shift somewhat between designs (most notably at 100\%,
where the interleaved mean is about half the sequential-block mean),
consistent with the modest time-drift detected above. But the
qualitative, security-relevant claim -- that the 40\% mid-range
condition is \emph{not} the low point of a monotonic decline, since
100\% sits above it in both designs -- reproduces under a design that
specifically controls for time confounds.

\begin{figure}[t]
\centering
\includegraphics[width=\linewidth]{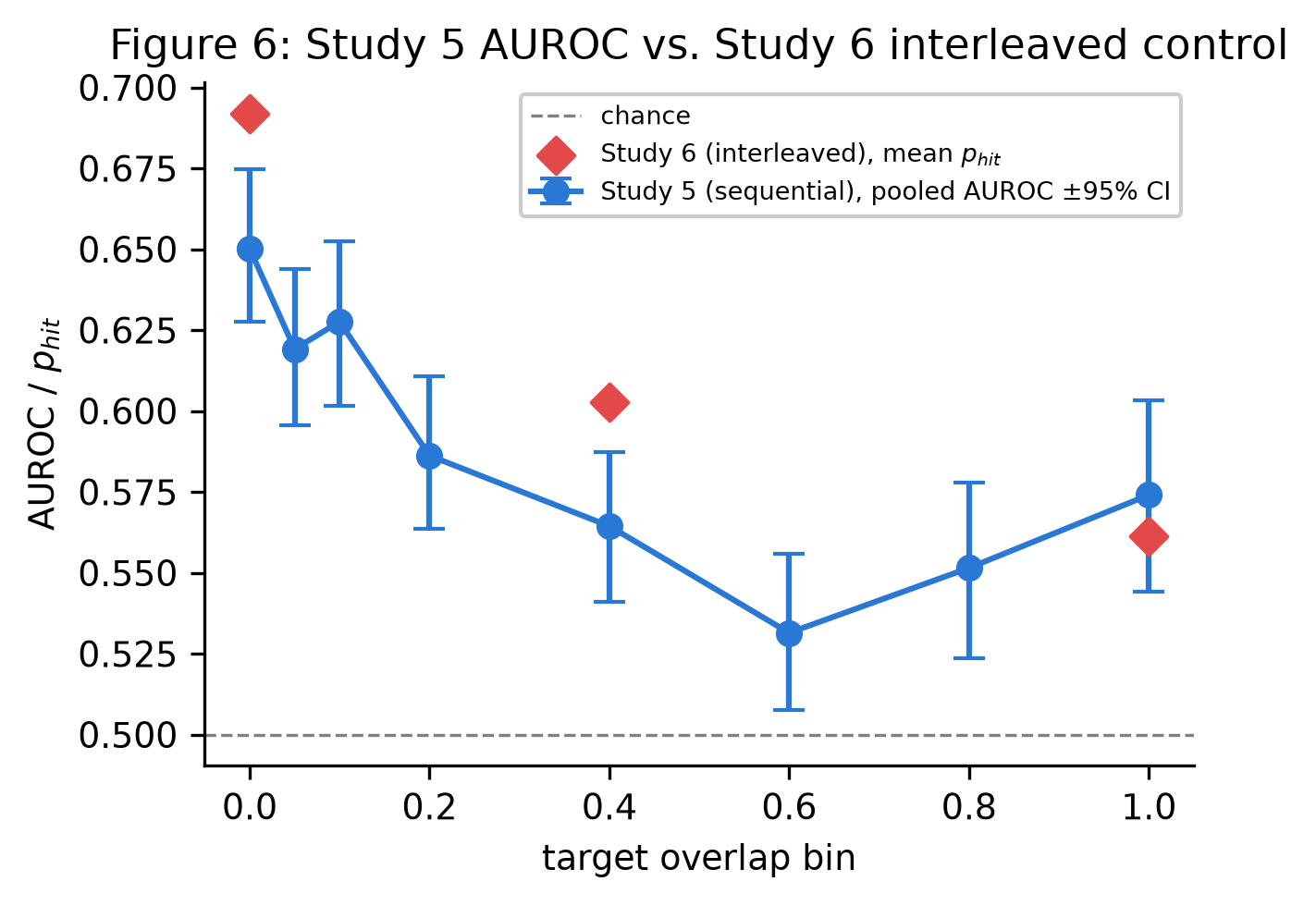}
\caption{Study 5's pooled AUROC per bin (blue, with 95\% CI) versus
Study 6's interleaved mean $p_{\mathrm{hit}}$ for the three re-run
conditions (red diamonds). The 40\% dip and partial 100\% recovery
appear in both the sequential and interleaved designs.}
\label{fig:auroc-interleaved}
\end{figure}

\subsection{Interpretation}

Study~6 supports treating the dip-then-partial-recovery pattern as more likely a real, if modest, non-monotonic feature than a sequential-block artifact. The round trends are descriptive rather than causal evidence of drift. The design interleaves conditions but does not counterbalance their fixed 0\%--40\%--100\% within-round order, so a short-timescale position effect remains possible.

\section{Study 7: Cross-Instance, Cross-Framework, and Multi-GPU Replication}
\label{sec:study7}

The first six studies are deliberately deep but centered on one primary vLLM/GB10 stack. We therefore ran a compact replication using the same paired probe logic while changing physical instance, framework, and deployment topology. The goal is external validation, not a second full dose-response campaign.

\begin{table*}[t]
\centering
\caption{Study 7: compact replication across physical instances, frameworks, and deployment topologies. All loaded pilots use $+4$ synthetic workers except the independent-node attempt, for which only the ambient result is retained. ``NS'' denotes no statistically significant ambient--loaded difference at the pilot sample size.}
\label{tab:study7}
\footnotesize
\setlength{\tabcolsep}{4pt}
\begin{tabular}{@{}lccccccl@{}}
\toprule
Deployment & Framework & GPUs & Ambient $d$ & Loaded $d$ & Welch $t$ & Inference \\
\midrule
Primary node, Study~1 & vLLM 0.28.0 & 1 & $1.126\pm0.269$ & $0.213\pm0.096$ & -- & collapse \\
Independent physical node & vLLM 0.27.1 & 1 & $\approx1.03$ & -- & -- & ambient replicated \\
MPS co-resident pilot & SGLang 0.5.17 & 1 & $0.073\pm0.104$ & $0.190\pm0.142$ & 1.50 (NS) & inconclusive \\
Two-node tensor parallel & SGLang 0.5.17 & 2 & $0.356\pm0.212$ & $0.468\pm0.173$ & 0.91 (NS) & inconclusive \\
Two-node tensor parallel & vLLM 0.26.0 & 2 & $3.418\pm1.433$ & $0.511\pm0.687$ & $\approx4.09$ & collapse, $p<0.01$ \\
\bottomrule
\end{tabular}
\end{table*}

An independent GB10 node first reproduced the ambient signal: five runs gave mean $d\approx1.03$ (range 0.77--1.41; hit consistency 80--96\%), close to the primary-node pilot. A courtesy loaded test on that third-party server was stopped when the service became unresponsive, so no causal or loaded-condition claim is drawn from that incident.

The strongest replication uses genuine two-node tensor parallelism over two physical GB10 GPUs with NCCL-coordinated model/KV state. On vLLM, ambient mean $d=3.418$ ($s=1.433$) falls to $0.511$ ($s=0.687$) under four background workers. Welch's test gives $t\approx4.09$, $\mathrm{df}\approx5.7$, $p<0.01$. Thus the central collapse survives a materially different deployment topology and vLLM release.

SGLang does not reproduce the same point-estimate direction: the one-GPU pilot changes from $d=0.073$ to $0.190$, and the distributed two-GPU pilot from $0.356$ to $0.468$; neither difference is significant. These are not evidence of an opposite SGLang effect. Both SGLang pilots share an important confound: SGLang required CUDA MPS and remained co-resident with the production vLLM workload, so its ``ambient'' condition was already heavily shared and is not comparable to the primary vLLM ambient baseline. We therefore treat cross-framework generality as unresolved, while the vLLM result now has independent-instance and distributed two-GPU support.

\section{Cross-Study Synthesis}
\label{sec:synthesis}

\subsection{Replication Across Experimental Designs}
The contention-onset effect is not a result of one parameter sweep. Studies~1--4 establish and instrument the ambient-to-loaded change across worker counts and filler shapes; Study~5 samples the missing sparse-overlap region, quantifies attack success, and derives concurrency depth; Study~6 checks the non-monotonic ordering under interleaving. Study~7 then moves beyond the primary instance: the ambient signal replicates on a second physical GB10 node, and the collapse remains statistically significant on a genuine two-node, two-GPU tensor-parallel vLLM deployment. The SGLang pilots are inconclusive under MPS co-residency and therefore bound, rather than resolve, framework generality.

\subsection{A Two-Regime Reliability Model}
The completed data motivate a simple conceptual model
\begin{equation}
d=\begin{cases}
d_A+\epsilon,&Z=0,\\
d_C+\eta(S),&Z=1,
\end{cases}
\end{equation}
where $Z$ denotes the presence of sustained synthetic contention, $d_A$ is the higher ambient-regime effect, $d_C$ the lower contended-regime mean, and $\eta(S)$ captures variation induced by serving state $S$. The experiments strongly support $d_A>d_C$. They do not identify a universal $d_C$ or a universal breakpoint, because both depend on the evaluated model, server configuration, accelerator, and ambient workload.

The Study~5 change-point result sharpens this interpretation. A single-line model explains little of the 120-run variation ($R^2=0.011$), whereas the best two-segment fit reaches $R^2=0.536$ with $\hat\tau=0$ and a bootstrap 95\% CI of $[0.000,0.113]$. Because the two segments are fitted independently, part of this gain comes from allowing a level shift, not only a slope change. The edge-touching breakpoint CI therefore supports an ambient-versus-loaded split rather than an interior physical threshold. Within the loaded region, effect size is weakly non-monotonic rather than progressively decreasing.

\subsection{Why a Smooth Utilization Model Is Unsupported}
A simple model such as $d=\beta_0+\beta_1O+\epsilon$ is not an adequate description of the data. In Study~4, most loaded runs already have $O\gtrsim0.76$, leaving little dynamic range. Study~5 fills the intermediate region, yet measured overlap remains only weakly correlated with effect size ($r=-0.106$), while concurrency-depth variance is substantially stronger ($r=-0.416$). The same high-overlap regime can therefore contain stable and unstable timing behavior. This is important methodologically: request overlap answers whether a probe shared wall-clock time with any synthetic request, but it discards how many requests were simultaneously active and how quickly that depth changed.

\section{Mechanistic Interpretation}
\label{sec:mechanism}

\subsection{Prefill Savings Versus Scheduler Variance}
Automatic prefix caching reduces redundant prefix computation. If all other system state were fixed, a cached prefix should reduce the latency component associated with prefill. Under continuous batching, however, all other state is not fixed. A probe may queue behind other requests, enter a batch with a different number of active sequences, or encounter a different mixture of prefill and decode work than the adjacent member of the same pair.

A useful decomposition is
\begin{equation}
L=L_{cache}+L_{sched}+L_{exec}+L_{host},
\end{equation}
where $L_{cache}$ denotes cache-dependent work, $L_{sched}$ scheduler and queueing delay, $L_{exec}$ accelerator execution effects, and $L_{host}$ client/server overhead. Prefix reuse primarily changes $L_{cache}$, whereas contention can change both the mean and variance of $L_{sched}$ and $L_{exec}$.

Suppose the cache optimization saves approximately $\delta$ for a fixed probe. In a simplified model,
\begin{equation}
|d|\propto \frac{|\delta|}{\sigma_L}.
\end{equation}
Thus a channel can remain physically present while becoming statistically weak if contention inflates $\sigma_L$. This interpretation is consistent with the observed collapse without requiring contention to disable prefix caching; the live prefix-cache counters confirm that hits continued to occur during the campaign.

\subsection{Pair-Asymmetry Effects}
The paired before/after design suppresses slow drift but remains sensitive to fast state changes. Let the uncached and cached members of pair $i$ execute at $t_1$ and $t_2$. If
\begin{equation}
S(t_1)\neq S(t_2),
\end{equation}
part of the measured pair difference reflects a scheduler-state mismatch rather than only cache state. A short bursty worker can repeatedly transition between active and inactive states at a cadence comparable with the probe sequence. Depending on alignment, one member of the pair can arrive during a queue transition while the other does not. This gives a plausible explanation for the unusually large Study~3 variance of \texttt{1w\_small16} ($s_d=0.7112$) compared with \texttt{2w\_tiny4} ($s_d=0.0727$), despite both occupying the same low-mean regime.

The data do not directly identify the scheduler event responsible. High-resolution server-side instrumentation of running/waiting sequence count, prefill/decode admission, and batch composition would be needed to establish causality. We therefore treat scheduler-state asymmetry as a mechanism hypothesis, not a demonstrated cause.

\subsection{Concurrency Depth as the Missing Request-Layer Variable}
For background intervals $I_j=[s_j,e_j]$, the instantaneous synthetic concurrency depth is
\begin{equation}
C(t)=\sum_j\mathbf{1}[s_j\le t<e_j].
\end{equation}
For each probe window $P_i$, we compute
\begin{align}
\bar C_i&=\frac{1}{|P_i|}\int_{P_i}C(t)\,dt,\\
C_i^{\max}&=\max_{t\in P_i}C(t),\\
V_i&=\mathrm{Var}_{t\in P_i}[C(t)].
\end{align}
The overlap metric is equivalent to integrating only $\mathbf{1}[C(t)>0]$. It maps $C(t)=1,2,3,\ldots$ to the same binary state and therefore discards both depth and burstiness.

Study~5 shows that $V_i$ is the strongest measured request-layer correlate of timing reliability: $r=-0.416$ with $d$ and $r=-0.637$ with hit consistency, compared with $r=-0.106$ and $-0.504$ for measured overlap. It also retains a residual correlation of $-0.329$ with $d$ after regressing out overlap. These results support the hypothesis that temporal variability in concurrent work is more informative than average overlap, but they remain correlational because uninstrumented ambient traffic and internal scheduler state are not observed.

\subsection{Why Reliability Partially Recovers at Saturation}
Study~5 produces a non-monotonic attack-reliability curve: AUROC declines from 0.650 at ambient to 0.531 near 61\% overlap, then rises to 0.574 at saturation. Study~6 reproduces the corresponding ordering in effect size under interleaving: $0\%>100\%>40\%$. A plausible explanation is that intermittent mid-range contention causes more scheduler-state transitions between paired probes, while a highly saturated stream can settle into a more stationary loaded regime. This would increase average load while reducing pair-to-pair state asymmetry. The concurrency-variance result is consistent with that interpretation, but the present measurements cannot distinguish scheduler stability from other sources such as host queueing or batch-composition effects.

\section{Security Implications}
\label{sec:security}

\subsection{Quiet-System Measurements Can Overstate Attack Reliability}
A side-channel evaluation on an otherwise quiet server can expose a much cleaner latency difference than an attacker experiences on a live shared service. On the evaluated stack, Study~2 reduces mean $d$ from 0.7789 in the ambient condition to 0.2109 with the first sustained synthetic load. Study~5 provides the attacker-level view: pooled AUROC falls from 0.650 at ambient toward chance (0.531) in the mid-overlap regime. Therefore, a single attack-success number is incomplete unless the serving-load regime is reported.

This does not mean previous cache-timing attacks are invalid. It means their operational reliability depends on the environment in which probing occurs. A result measured during a quiet maintenance window and a result measured during a heavily multiplexed production period can both be correct while implying different probe budgets and error rates.

\subsection{Contention Is Not a Defense}
Contention should not be treated as a security boundary. It is uncontrolled, workload dependent, and may disappear during low-demand periods. An adaptive attacker can repeat probes, select favorable time windows, or build a classifier conditioned on observable load. Artificially creating background work would also consume capacity and degrade legitimate users. The appropriate defense remains cache isolation or a sharing policy that prevents untrusted principals from obtaining cross-tenant cache-hit signals.

Our results instead affect risk assessment. Operators evaluating a shared-prefix design should test the channel under multiple realistic load regimes, and attackers should be evaluated by the number of probes needed to reach a target decision confidence rather than by a single idealized latency gap.

\subsection{Probe Budget and Residual Exploitability}
A weak single-probe effect does not imply that inference is impossible. Study~5 explicitly evaluates repeated majority voting. For ambient traffic, $p_{hit}=0.695$ yields $P_{succ}(9)=0.895$. Near 61\% measured overlap, $p_{hit}=0.525$ produces only $P_{succ}(9)=0.562$. At saturation, $p_{hit}=0.584$ yields $P_{succ}(9)=0.699$. Thus contention materially increases the sample complexity of a simple attacker, but the channel remains above chance in every tested bin.

The analysis uses odd $k$ so that no tie-breaking rule affects the result. It is intentionally conservative in another sense: majority vote uses only the sign of each paired latency difference and discards its magnitude. A stronger attacker could aggregate continuous scores or condition on observable service state. Consequently, the reported probe-budget curve should be read as a characterization of a simple repeated-probe attacker, not an upper bound on exploitability.

\subsection{Reporting Guidance}
Cache-timing evaluations should report concurrent-request structure, prompt/output lengths, prefix-cache state, scheduler configuration, latency scope, probe budget, serving framework/topology, and whether background load is controlled or ambient. Run-level distributions are preferable to means alone.

\section{Validity, Limitations, and Reproducibility}
\label{sec:validity}

\subsection{Internal Validity}
The largest internal-validity threat is uncontrolled ambient production traffic. We intentionally retain it because the research question concerns a live shared server, but it adds variance that a fully isolated testbed would remove. The zero-synthetic-worker condition therefore means zero experiment-generated filler traffic, not zero total contention. Study~6 reduces long-timescale sequencing bias by interleaving 0\%, 40\%, and 100\% conditions round-by-round. It does not fully counterbalance them: every round uses the same within-round order, so a short-timescale position effect cannot be ruled out.

Studies~1--5 also use sequential condition blocks. Slow changes in ambient activity could therefore contribute to between-condition differences. The interleaved replication shows that the key $0\%>100\%>40\%$ ordering survives a different schedule, which reduces but does not eliminate this concern. A future replication could randomize all eight Study~5 bins within each block.

\subsection{Construct Validity}
Worker count is a configuration knob, not a physical GPU-load measure. Generation length similarly changes work duration without uniquely determining instantaneous accelerator pressure. Study~4 therefore moves to measured request overlap, but that quantity is still a request-layer temporal coverage metric rather than SM occupancy, CUDA kernel occupancy, memory-controller utilization, or vLLM batch occupancy. Concurrency depth improves the request-layer representation but still does not expose internal scheduler state.

Timings are end-to-end non-streaming request completion rather than direct server-side TTFT. Because attack probes request one output token, the measurement is dominated by request handling, prefill, and one-token decode, but it is not identical to TTFT. This boundary matters when comparing our absolute latencies with work that instruments TTFT directly.

\subsection{Statistical Conclusion Validity}
Studies~1 and~4 have small per-condition sample counts and are used primarily for hypothesis formation and instrumentation. Study~2 uses 12 runs per worker level and reports bootstrap CIs plus planned Welch contrasts. Study~5 is the largest experiment with 120 runs, 750 paired trials per overlap bin, bootstrapped AUROC CIs, and a bootstrapped change-point analysis. Failure to detect a difference between two higher-load worker levels is not evidence that their distributions are equal; our wording is restricted to ``no statistically detectable additional loss.''

The segmented model is intentionally unconstrained across the breakpoint. This allows a discontinuous level shift and therefore can improve $R^2$ through both intercept and slope changes. We report the model exactly as implemented and do not interpret its $R^2$ gain as evidence of a smooth physical kink. Likewise, Pearson correlations are descriptive associations, not causal estimates.

\subsection{External Validity}
The experiments still use one model family and one accelerator architecture (GB10), so numerical effect sizes and scheduler dynamics remain stack specific. However, Study~7 expands the evidence to two physical nodes, three vLLM releases/configurations, one- and two-GPU deployments, and SGLang 0.5.17. The central vLLM collapse survives genuine cross-node tensor parallelism, reducing the single-instance concern. Framework generality is not established because the SGLang pilots are statistically inconclusive and their MPS co-residency prevents an equivalent low-contention baseline. Replication on a different GPU architecture and an independently idle SGLang host remain the most valuable external-validity tests.
Scripts and raw data have since been archived, with internal network
addresses redacted, at:
\url{https://github.com/engranaabubakar/kv-cache-contention-timing-sidechannel}.

\subsection{Reproducibility and Artifact Scope}
The primary server configuration is fully specified for directly observed variables: vLLM 0.28.0, PyTorch 2.13.0+cu130, CUDA 13.0, NVIDIA driver 580.95.05, half precision, GPU-memory-utilization 0.70, maximum model length 4096, eager execution, and one accelerator. Study~7 additionally records the vLLM 0.26.0/0.27.1 and SGLang 0.5.17 replication configurations, tensor-parallel topology, and MPS status. Prefix-cache activity was verified from the live metrics counters rather than inferred from a default configuration.

The experiment records raw paired timings, per-run $d$ and $H$, probe/background start and end times, achieved overlap, and configuration metadata. The artifact contains the probe, instrumented background generator, sparse-contention generator, concurrency-depth analysis, attack-reliability analysis, change-point analysis, and interleaved-control scripts. Because the interval representation is retained, overlap and all concurrency-depth summaries can be recomputed deterministically from the original logs.

\section{Discussion}
\label{sec:discussion}

\subsection{What the Data Establish}
The strongest supported conclusion is that cache-timing reliability on vLLM is highly sensitive to the onset and temporal structure of concurrent activity. Studies~2--6 establish the regime change, reject an interior overlap threshold, and identify concurrency-depth variance as the strongest request-layer correlate. Study~7 strengthens external validity: the collapse reproduces significantly on distributed two-GPU vLLM, while SGLang remains unresolved rather than contradicting the result. Together, these findings are stronger than the generic statement that ``load adds noise.''

\subsection{What the Data Do Not Establish}
The data do not establish a universal threshold, that timing attacks become impossible, or that all serving frameworks respond identically. They also do not identify a specific GPU or scheduler mechanism. Concurrency-depth variance is the best measured correlate, but the link is correlational and ambient production work is not instrumented. In particular, the SGLang pilots cannot support an opposite-effect claim because their baseline is MPS co-resident and their ambient--loaded differences are not significant.

\subsection{Operational Interpretation}
For an operator, the result means that measured side-channel severity can depend strongly on when and how testing is performed. A test during a quiet period may observe a much cleaner timing signal than normal multi-user operation; a noisy daytime measurement, conversely, should not be interpreted as proof that cache sharing is safe. The vulnerability is created by cross-principal reuse, while contention changes the attacker's observation quality.

\section{Conclusion}
This paper characterizes KV-cache timing-side-channel reliability under multi-tenant contention. Across seven studies, the primary vLLM stack shows an ambient-to-loaded reliability collapse rather than a gradual worker-count response. A 120-run sparse-overlap experiment finds no interior physical threshold; attack-level AUROC approaches chance in the mid-overlap regime, while concurrency-depth variance is the strongest measured correlate of residual instability. An interleaved control preserves the non-monotonic ordering. Crucially, the central collapse also reproduces on a genuine two-node, two-GPU tensor-parallel vLLM deployment ($3.418\rightarrow0.511$, $p<0.01$), reducing the single-instance concern. SGLang pilots are statistically inconclusive under an MPS co-residency confound, so framework generality remains open. The broader lesson is methodological: side-channel severity measured on a quiet serving stack should not be assumed to represent attack reliability under concurrent production load, and serving framework/topology must be reported as part of the security context.

\bibliographystyle{IEEEtran}
\bibliography{kv_cache_contention_refs}

@inproceedings{wu2025promptpeek,
  author    = {Guanlong Wu and Zheng Zhang and Yao Zhang and Weili Wang and
               Jianyu Niu and Ye Wu and Yinqian Zhang},
  title     = {I Know What You Asked: Prompt Leakage via KV-Cache Sharing in
               Multi-Tenant LLM Serving},
  booktitle = {Network and Distributed System Security Symposium (NDSS)},
  year      = {2025}
}

@article{song2024earlybird,
  author  = {Linke Song and Zixuan Pang and Wenhao Wang and Zihao Wang and
             XiaoFeng Wang and Hongbo Chen and Wei Song and Yier Jin and
             Dan Meng and Rui Hou},
  title   = {The Early Bird Catches the Leak: Unveiling Timing Side Channels
             in LLM Serving Systems},
  journal = {arXiv preprint arXiv:2409.20002},
  year    = {2024}
}

@article{zheng2024inputsnatch,
  author  = {Xinyao Zheng and Husheng Han and Shangyi Shi and Qiyan Fang and
             Zidong Du and Xing Hu and Qi Guo},
  title   = {InputSnatch: Stealing Input in LLM Services via Timing
             Side-Channel Attacks},
  journal = {arXiv preprint arXiv:2411.18191},
  year    = {2024}
}

@article{chu2026safekv,
  author  = {Kexin Chu and Zecheng Lin and Dawei Xiang and Zixu Shen and
             Jianchang Su and Cheng Chu and Yiwei Yang and Wenhui Zhang and
             Wenfei Wu and Wei Zhang},
  title   = {Selective KV-Cache Sharing to Mitigate Timing Side-Channels in
             LLM Inference},
  journal = {arXiv preprint arXiv:2508.08438},
  year    = {2026},
  note    = {Version 2, revised February 2026}
}

@article{addagada2026kvgov,
  author  = {Tejasvi C. Addagada},
  title   = {Governing the KV Cache: Preventing Timing Side-Channel Leakage in
             Multi-Tenant LLM Inference},
  journal = {arXiv preprint arXiv:2608.09225},
  year    = {2026}
}

@inproceedings{kwon2023pagedattention,
  author    = {Woosuk Kwon and Zhuohan Li and Siyuan Zhuang and Ying Sheng and
               Lianmin Zheng and Cody Hao Yu and Joseph Gonzalez and
               Hao Zhang and Ion Stoica},
  title     = {Efficient Memory Management for Large Language Model Serving
               with PagedAttention},
  booktitle = {Proceedings of the 29th Symposium on Operating Systems
               Principles (SOSP)},
  year      = {2023}
}

@misc{vllmapc2026,
  author       = {{vLLM Project}},
  title        = {Automatic Prefix Caching},
  year         = {2026},
  howpublished = {\url{https://docs.vllm.ai/}},
  note         = {Accessed September 2026}
}

@misc{vllmmetrics2026,
  author       = {{vLLM Project}},
  title        = {vLLM Metrics},
  year         = {2026},
  howpublished = {\url{https://docs.vllm.ai/}},
  note         = {Accessed September 2026}
}
\end{document}